\documentclass[apjl,a4paper,12pt,useAMS]{emulateapj}
\usepackage{amssymb}
\usepackage{graphicx}

\begin{document}
\title{An Excessive Core Collapse in Nbody Cosmological simulations}

\author{Weike Xiao\altaffilmark{1},Xufen Wu\altaffilmark{2}}

\altaffiltext{1}{Department of Astronautics Engineering, Harbin
Institute of Technology, POB 333, Harbin, 150001, China,{wkxiao@hit.edu.cn}}
\altaffiltext{2}{Department of Astronomy, University of Science and Technology of China,
Hefei, 230026, China}

\begin{abstract}

The particle mass used in cosmology N-body simulations is close to
$10^{8}M_{\odot}$, which is about $10^{65}$ times larger than the
GeV scale expected in particle physics. However, self-gravity
interacting particle systems made up of different particle number and mass
have different statistical and dynamical properties. Here we
demonstrate that, due to this particle number and mass difference, the nowaday
cosmology N-body simulations can have introduced an excessive core
collapse process, especially for the small halos at high
redshift. Such dynamical effect introduces an excessive cuspy center
for these small halos, and it implies a possible connection to the
so called "small scale crisis" for CDM models. Our results show that
there exist a physical limit in cosmological simulations, by using about
 $10^3$ particles to describe smallest halos, and we provide a simple
suggestion based on it to relieve those effects from the bias.

\end{abstract}

\maketitle

\section{Introduction}


With  the rapid development of
computer science, the N-body cosmology simulation has become an
important method for studying dark matter particle systems. Such
numerical experiment method shows the Cold Dark Matter(CDM)
universe with a dark energy parameter $\Lambda$ can
have nice agreement with observations on large scale topics (Kuhlen,M.et al.2012,Springel V.et al.2006).
But the numerical predictions on small scale topics depart from the
observations: High-resolution rotation curves of low surface
brightness galaxies show the halo density profiles have flat cores
(Burkert 1995,de Blok 2002,de Blok 2005,Gentile 2005),
yet the simulation results tell us they should have cuspy
centers (Navarro 1997,Navarro 2004)
Simulation results also predict about 10 to 100 times more subgalaxies round our Galaxy
(Klypin 1999,Moore 1999a ,Springel 2008) and the subhalos are too dense(see Boylan 2011).
This is the so called "small scale crisis" and caused people's
suspicion on CDM models.

Many different explanations were carried out to displace the
traditional CDM models, such as the Warm Dark Matter(WDM)(Colombi 1996),
the Self Interacting Dark Matter (SIDM)(see Spergel 2000,Dave 2001),
the MOdified Newtonian Dynamics (MOND)(Milgrom 1983)
models and effect of baryons(Governato 2010).
Yet the new models have also caused new disputes for themselves
(e.g. see Yoshida 2003,Markevitch 2004,Zhao 2006,Kuzio 2010).

Anyway, on small scales topics, the numerical simulation results are also the
mainly basic causing the suspicion of the CDM. Since the numerical method are
still not perfect, such as the limited particle number, the limited time step,
the dynamic of no time delay system etc., the bias in the simulation results can
also cause the CDM small scale problems, and we should discuss in which term do
the numerical results become dependable computing.

\section{Physical Bias}
Here we notice that there exist a
physical bias in nowaday simulations: the particle number and particle mass.
Due to the technical limitation, the particle numbers used in cosmological
simulations are limited. To obtain the mean density of the universe
, people have to set a huge mass for each particle in simulations.

With the improvement of computer science, the particle numbers used
in simulations have increased from $~10^6$ to about $~10^{11}$ within
the last decade, and the particle mass used on small scale topic has
decreased from $~10^{10}M_{\odot}$ to $~10^{3}M_{\odot}$ (Springel 2008)
. It is acceptable to set the particle mass as
$10^{11}M_{\odot}$ when studying the evolution of large scale
structure, for we can explain each particle as one galaxy. But for
the small scale topics, such as the dark matter halo property, the
galaxy formation, the galaxy merging process, the first star
formation and etc., the simulated particle mass is still about a
factor of $f \sim 10^{65}$ times larger than the expected $100GeV$
candidates in particle physics(Gaitskell 2004).

At the same time, the particle number density $n_{sim}$ also has the same factor
smaller than expected $n_{DM}$ (that means using only about $10^{1} \sim
10^{8}$ particles to simulate one $~10^{13} M_{\odot}$ dark matter
halo). If we don't think the mass of one dark matter particle can be
heavier than our human body, we should consider that a physical bias
exist in simulations:
\begin{eqnarray}
&m_{sim}&\rightarrow m_{DM}\times f, \nonumber\\
&n_{sim}&\rightarrow n_{DM}/f
\end{eqnarray}
Do these two kinds of self-gravitating particle systems with such a
bias have the same statistical and dynamical properties? If not, one
should be cautious when applying the simulation results.

\section{Long Term Relaxation }
Generally, particles in a
gravitational potential $\Phi(\emph{r})$ follow its equation of
motion:
$$\ddot{\emph{r}}_i = -\nabla\Phi$$

The equation does not include particle mass $m_i$, one might expect
particles with different mass (even with a bias of $10^{65}$) can
follow the same orbit, just like Galileo's two balls of different
weights. However, remember the potential of the system
$\Phi(\emph{r})=\Sigma\Phi_i(r)$ is assembled by potential of each
member, the bias means a different assembling method.

It is not easy to describe the difference from the bias in general.
But for a virialized dark matter halo, the numerous theoretical and
numerical studies on globular clusters can give us much help. One
point is the long term relaxation effect.

From the view point of one particle, when it is flying in a stable
and spherical halo potential, theoretically its energy $E_i$ and
angular moment $\bf{L}_i$ should be conserved as constants. But for
a virialized dark matter halo, the potential is contributed by many
moving particles, that means the potential will no longer be ideally
spherical and stable(It might be acceptable to describe such a
stable and spherical system using about $10^{70}$ particles, but
hard to be accepted by only $10^5$ or even $10^1$ particles). In
this case, both $E_i$ and $\bf{L}_i$ can be changed. Intuitively,
when the halos include fewer particles, such effect will be more
serious. Note that such long term relaxation effect we mention is
caused by particle density field fluctuations on large distance, but
not by collisions of a few close particles (this is another way
changing $E_i$ and $\bf{L}_i$).

Galatic Dynamics (Binney GD 2008) show the relaxation time scale
$t_{relax}$ caused by long term particle density field fluctuations
(here the softening parameter does not affect the results) should
be:
\begin{equation}
\label{trelax} t_{relax}\simeq   (0.1 N/ln N) t_{cross}
\end{equation}
where $N=M/m_i$ is the particle number of the halo, $t_{cross}=R/v$
is the crossing time scale and $v$ is the virial velocity
($v^2\simeq GM/R$) . Analytical and simulated results (see Huang 1993,Diemand 2004)
 give the similar formula. For one halo with given $M$:
\begin{equation}
\label{trelax_f} t_{relax}\propto ln f/f
\end{equation}
 The
Eq.(\ref{trelax_f}) shows us that the bias greatly shortens the
relaxation time.

If we define the mean free path as $L_s\equiv v t_{relax}$, then we
can follow the SIDM models (Spergel 2000,Dave 2001)
defining the "scattering cross section" as $\sigma\equiv
1/(L\rho)$. In the central region of a typical simulation halo, the
scattering cross section is about $\sigma_{sim}\simeq 9\times
10^{-26}cm^2/GeV$ (Xiao 2004), that is approximately the value
expected in SIDM models ($\sigma_{sim}\simeq 0.1\sigma_{SIDM}$). In
contrast, for the $GeV$ CDM particles $\sigma_{CDM}\simeq
10^{-65}\sigma_{SIDM}\simeq 0$.

Now we find the difference: the bias has bring an {\it excessive}
scattering cross section for the CDM models. The value of
$\sigma_{sim}$ cannot be neglected for CDM models, but not big
enough for SIDM models. Will it affect the dynamical properties of
the halos?

\section{Core Collapse }
The excessive scattering cross
section means particles in simulations will have an excessive way to
exchange their energy and angular momenta. Then the simulation halos
are possible to follow the evaporation effect appearing in globular
clusters: Once a particle exchanges its energy and gets $E_i>0$, it
can fly away and never come back. In a virialized system, the mean
particle energy $<E_i>=-GM/R<0$. That means the evaporating
particles always bring out energy, and the left particle system
becomes tighter and tighter. Such process appear more serious at the
central part of the halo, and the result is to introduce a dynamical
core collapse of the system.

Such evaporation and core collapse processes have been well studied
in galactic dynamics on the topics of stellar clusters. Since the
dark matter halos in nowaday simulations are similar to the globular
clusters: both are virialized systems and consisted of pure
gravitational interacting particles, and even have the similar
particle numbers (about $10^1$ to $10^8$); we can use the same
method to estimate their core collapse time scales. Following the
way analyzing stellar clusters (see Spitzer 1969,Giersz 1994,Binney GD 2008)
we get the core collapse time scale of a virialized dark matter halo in
simulations with $t_{rh}=\frac{0.17N_{halo}}{ln(0.1N_{halo})} \sqrt{\frac{{r_h}^3}{GM}}$,
or rewrite it as:
\begin{equation}
\label{tcc}
t_{cc}\simeq t_u\frac{0.003}{1+z} \frac{N_{halo}}{ln(0.1N_{halo})}(\frac{M}{10^{12}M_\odot})^{-\frac{1}{2}}(\frac{r_h}{10kpc})^{\frac{3}{2}}
\end{equation}

Here $M$ is the halo mass, and $N_{halo}$ is the particle number of the halo, $r_h$ is
the half mass radius, $t_u=1.37\times10^{10}yr$ is the Hubble time
in $\wedge$CDM models, and we have suggested $t_{cc}$ to be about 16
times of the half mass relaxation time $t_{rh}$ (see Takahashi 1995).

Before discussing in detail, we should emphasize the effect of the
softening parameter $\epsilon$ introduced in simulations. Softening
is a numerical trick introduced in N-body simulations to prevent
numerical divergences when two particles become very close (and the
force goes to infinity), the method is to modify each particle
gravitational potential, such as the form
$\Phi=-\frac{1}{\sqrt{r^2+\epsilon^2}}$. The introduction of
$\epsilon$ can effectively affect the short term "two-body
relaxation" process. However,

$(1)$ The softening parameter $\epsilon$ is unable to make the halos
avoid such core collapses.  Because the gravitation is a long term
interaction, the relaxation process discussed above is mainly caused
by the long tern particle encounters. The introduction of $\epsilon$
has no business with these long term process. In fact, the time
scale derivation of eq.(\ref{trelax}) and eq.(\ref{tcc}) in Galatic
Dynamics is based on the discussion of the density field
fluctuations in distance and $\epsilon$ will not change it.

$(2)$ One other point is that $\epsilon$ prevents the hard binaries
formation. The hard binaries release energy and drive a reexpansion
of the core after the core collapse in a globular cluster (e.g. Cohn 1989),
yet the softening parameter $\epsilon$ makes such
processes impossible for dark matter halos in a cosmological
simulation.

\section{A Physical limitation }
Equation (\ref{tcc}) can
tell us many secrets. For the N-body cosmological simulation process,
we focus on the dynamical property of visualized spherical halos.

First, comparing different resolution simulations for a given dark matter
halo (with setting value of $M$ and $r_h$),we find the core collapse time scale is
proportional to the particle number of the halo: $t_{cc} \propto N_{halo}/ln(0.1N_{halo})$.
For the $GeV$ CDM particle halo, $t_{cc}\gg t_u$ and
the core collapse will never happen within one Hubble time. But for
one Galaxy dark matter halo in simulations, if we use less than
$N_{halo}\simeq 10^{12}M_\odot / (10^9M_\odot) \simeq 1000$ particles
to progress the simulations, the bias of particle mass will bring an
excessive core collapse within one Hubble time. Our result shows a
limitation of the particle numbers $\sim10^3$ that should be used
when studying the Galactic scale topic in simulation.

Second, comparing different halos in one simulation, since $m$ is setting,
equation (\ref{tcc}) shows us $t_{cc}\propto M^{\frac{1}{2}}r_h^{\frac{3}{2}}$,
this means the $t_{cc}$ are longer for larger halos which should have larger
$M$ and $r_h$. Or to say, smaller halos are more dangerous.

Since the $\Lambda CDM$ models show us a hierarchical structure
formation scenario, the most dangerous halos are the "leaves of the merger tree".
we expect to avoid such core collapse process in the whole cosmological simulation,
if we ensure all the smallest halos at the beginning follows the limit:
\begin{equation}
\label{lim} t_{cc} \geq \alpha t_u
\end{equation}
The $\alpha t_u$($\alpha \leq1$) is the mean time scale of these
smallest halos existing in the universe before merging. If we set
$\alpha=1$, then we ensure the core collapse process caused by the
relaxation effect will not happen in the smallest halos (so for all
the larger halos within the whole hubble time).

The parameter of the smallest halos at high redshift are decided by
the initial conditions of the simulations. In nowaday cosmological
simulations, people apply the linear theory and use Fourier power
spectrum $P(k)$ to describe the initial fluctuations $\delta(x)$,
and to generate the initial conditions(see Seljak 1996,Springel 2008,
GRAFIC2 Bertschinger 2001).
 But due to numerical limitation, the simulation initial conditions can
only represent part of $P(k)$ in a limited range
$[k_{min},k_{max}]$, where $k_{min}$ is decided by the simulation
box size, and $k_{max}$ figures the smallest halo properties at the
beginning.

In hierarchical structure formation scenario, the smallest halo
formed by the collapsing of the dark matter within one shortest
wavelet $\lambda=2\pi/k_{max}$. So we estimate the mass of it as:
$M\simeq\bar{\rho}\frac{4\pi}{3}(\frac{\lambda}{2})^3
\simeq\bar{\rho}\frac{130}{k_{max}^3}$. In halo models (see ShaunCole 1996)
the spherical collapse halos have the mean density of about
$178\bar{\rho}$, then their characteristic radius $r_{0}$ follow
$178\bar{\rho}r_0^3\sim\bar{\rho}(\frac{\lambda}{2})^3$, if we set
$r_h\sim0.1a r_0$, (for NFW density profile $2 \leq a\leq3$ ), then
we get:
$r_h\simeq0.1\frac{a}{\sqrt[3]{178}}\frac{\lambda}{2}\simeq\frac{5.58\times10^{-2}a}{k_{max}}$.

Combining eq.(\ref{lim}) and eq.(\ref{tcc}), we find the limit:
(here $\rho_0\equiv1.4\times10^{11}M_\odot/Mpc^3$ for a $\lambda$CDM model)

\begin{equation}
\label{Nmin}
\frac{N_{halo}}{ln0.1N_{halo}}>600\alpha (1+z)(\rho_0/\bar{\rho})^\frac{1}{2}
\end{equation}

The result is interesting, it is not sensitive with $k_{max}$, that means
no matter cosmological simulation of what kind of scale, the limitation is the same:
people should use enough particles to describe the smallest halo. The discussion
of $\alpha(1+z)$ may be complex but it is setting in one simulation. For a
$\lambda CDM$ Universe, if we believe $\alpha(1+z)\simeq10^0$, the solution
of eq.(\ref{Nmin}) is about $N_*\simeq 3500$.

Wether the Virialized dark matter halos show us the dynamical difference within one Hubble
time, depends on the resolution of the them. When we use enough particles as $N_{halo}>N_*$
for the smallest halo in the simulation, we can ensure all halos avoid the core collapse
process caused by the unexpected relaxation effect. But if one use too few particle to
describe these smallest halos, the hugely magnified scattering cross section can introdue
an excessive core collapse for all the smallest halos, and the cores do not re-expand like
a globular cluster due to the softening parameter.  The dynamical difference between the
"ant particle" ($m\sim GeV$) system and "elephant particle" ($m\sim 10^{8} M_\odot$) halo
will be serious in this case.

\section{An Excessive Core Collapse }
Though the limitation given by eq.(\ref{Nmin}) is just a very rough estimation, but we can
see the particle number should be on the scale of $10^3$ to avoid such core collapse for
the smallest halo. This situation must be satisfied for one familiar
$M\simeq10^{10}M_\odot$ halo containing GeV CDM particles in nature.

Then how about the situation in nowaday cosmological simulations? Though the limitation seems
possible to fulfil, how ever, we find people are always trying to set the $k_{max}$ to be the
Nyquist frequency when generating the initial condiction in nowaday cosmological Nbody
simulations(see Springel2005,Scoccimarro 2012). Yet the Nyquist frequency of one
dimension means "only two points within one wavelength", then the smallest three dimensional
structure can only include less than $2^3=8$ particles, so one can imagine those "smallest halo"
will be impossible to have $N_{halo}$ more than the $N_*$ limitation. Or to say, such
simulations have surely introduced an excessive core collapse for their smallest halos.

One might believe the recent re-simulation methods (see Hahn 2011,Springel 2008)
can solve the problem by introducing much more particles (about $10^3$ times more) to the same region.
Unfortunately, when generating the initial conditions to represent
$P(k)$ in [$k_{min}$,$k_{max}$] in the resimulation region, people
are still artificially setting $k_{max}$ to be the new technical
limitation $k_{ny}'$. Therefore, in these simulations $k_{max}'=k_{ny}'$,
no matter how large the particle number $N$ is, and the smallest halos are always
containing about 8 particles, the limitation in eq.(\ref{Nmin}) is
always violated. Hence we find these simulations have definitely
introduced an excessive core collapse for their high redshift small halos.

\section{Fossils in the Simulation }
Though we have
demonstrated the unavoidable existence of the excessive core
collapse process for the small halos qualitatively, it is not easy
to make clear how will it tamper with the simulation results
quantitatively. However, the dynamical effect of the high redshift
small halos is apparent: the excessive core collapse (maybe more
than one times for the same halo) can make the halos to be more
concentrated than they should be, and makes their density profiles
to be much more cuspy at central regions.

In a hierarchial structure formation scenario, these high redshift
small halos will soon bring themselves into complex merger process:
they merge into each other or are devoured by huge halos. A lot of
theoretical and numerical studies have been carried out on the
merger process and shown us the qualitative properties:

$\bullet$ Major Merger process: When two halos with similar mass
merge together, theoretical study(Dehnen 2005) and numerical
experiments (Michael 2003,Ileana 2008) showed us the two
halos syncretize together, the center of one halo sinks rapidly to
the center of the other halo. But the central density sloop
information can be well retained: a major merger of two-cored halos
yields a one-cored halo; yet mergers between a cuspy halo and a
core/cuspy halo, the inner density sloop of the remnant will be
closer to that of the steeper one of the initial systems.

$\bullet$ Minor Merger process: For the merger process between a
large halo and a satellite ($M_s \leq 0.1 M_h$), semi-analytical and
N-body method study(Taffoni 2003) show us the fate of
the satellite halo is determined by its orbit and concentration
property: low concentration satellite below $0.1M_h$ is disrupted by
tides quickly; yet high concentration one can survive with a low
mass center and become a new substructure of the large halo.

Since the dark matter halos are assembled step by step from the high
redshift small halos in hierarchical CDM halo models, one can image
how the excessive core collapse affects the simulation results:

$(0)$ It make the earlier small halos to be too concentrated and
leave them a much too steep density sloop at the central region.

$(1)$ For a halo mainly experience major mergers, each merger process
remain the center character of more cuspy member. Retaining such process
later on, including the final product at $z=0$ will get a too cuspy
center (causing the "cusp problem"?).

$(2)$ For a halo mainly experience minor mergers, the too concentrated
center means it will have too high survival probability from the merger
and left an "excessive substructure" (causing the "substructure problem"?).

$(3)$ For a halo experience major and minor merger alternately, it can
survive as an independent subhalo but keeping the too cuspy property
(causing the "Too big to fail problem"?).

So our discussion imply a possible connection with the unexpect core
collapse and the three "small scale crisis" problem of CDM.

It is possible that some other physical mechanisms bring about the main
unconformity between simulation halo properties and observations, then
"causing the problem" just changed as "amplify the problem". Yet since
our discussions have shown the excessive core collapse can bring unwanted
fossils, when discussing other possible mechanisms, people should avoid these
unwanted fossils to get reliable simulation result.

\section{Can we Avoid it? }
Maybe people are just trying to
represent more information of the power spectra $P(k)$ when generating
the initial condition for the simulation. But setting $K_{max}=k_{ny}$
have introduced some too high frequency information of $P(k)$ that are
described with not enough particles.

Considering the smallest halo follow $N_{halo}\propto\lambda_{min}^3\propto (k_{max})^{-3}$,
one can also define a "safety frequency" with $N_*$:
\begin{equation}
\label{Kmax}
k_*=k_{ny}(\frac{8}{N_*})^\frac{1}{3}
\end{equation}

One simple property of $k_*$ is that $k_*<k_{ny}$, which means the
smallest halo should NOT be described with only about 8 particles.

It is the small halos corresponding to the wavelength between $k_*$ and $k_{ny}$ who
introduced these fossils. The reason is that the traditional method
can not give enough particle numbers to model the halos on this
scale. Hence we suggest, we should ensure the physical limitation $k_{max}\leq k_*$
when setting the initial conditions, but not use the technical limitation
$k_{max}=k_{ny}$.

One excessive subhalo relate to the fossil within one high
redshift small halo, yet the too cuspy density profile of a huge halo
correspond to all fossils within each small halos of its merger
tree. We can expect when people use $k_* < k_{max} < k_{ny}$ in
simulation, they can see the subhalo number decrease serious but the
too cuspy density profile will not change unless they use $ k_{max}
< k_*$.

\section{Numerical experiment scenario }

If our suspects are
correct, one can also predict a simple "numerical experiment scenario":
We can repeat one $512^3$ particle CDM simulation about ten years ago
(with $k_{max}=k_{ny}$), then we set the box size $L$ and $k_{max}$
unchanged, but using $1024^3$ and $4096^3$ particles to do the same
simulation. Then we can see only the "substructure problem" can be
released in $1024^3$ case; both the substructure and cuspy problems
can be released in $4096^3$ case (if $N_*\simeq 3500$). Then we can
change $k_{max}$ to be the new $k_{ny}'$ of $4096^3$ case, the new
simulation result can show us both problems again.

Some numerical experiments have already shown us interesting
results. Actually, many nowaday WDM simulations can be considered
as the numerical experiments for us: their main difference with CDM simulations
is cutting down the high frequency part of $P(k)$, something like setting
$k_{max}<k_*$ in CDM case. Their results show both the "cusp problem" and
"substructure problem" can be released seriously.

Moore et al.(1999b) have compared results with limited
$k'_{max}$, which is less than $k_{ny}$ (though still larger than
$k_*$) and the normal case ($k_{max}=k_{ny}$). Their results show
the huge halo still contain too cuspy density profile (see figure3),
but the number of substructures dropped seriously (see figure4). We
can now expect the too cuspy density profile can also be changed
when they use $ k'_{max} < k_*$.

Since the dangerous small halos correspond to wavelength of
$k_*<k<k_{ny}$, the traditional method simulation (setting
$k_{max}=k_{ny}$) with higher resolution ($k_{ny}$ will be larger,
corresponding to larger number of smaller halos), will show us a
much more larger number of smaller subhalos in the simulation
result. Just as shown in Via Lactea simulation (e.g. Kuhlen M. et
al. 2008). And we predict similar result can
still be seen if Kuhlen M. use more higher resolution.

Another point is that $N_*$ in eq.(\ref{Nmin}) can help us avoid the
excessive core collapse, but not every thing from this bias. Some
other topics, for instance, the translation of angular momentum,
should be studied further in details.


\section{Discussion }
In summery, we discussed a long term
relaxation effect which is amplified hugely by the bias of particle
mass/number in cosmological simulations. With the mature theories
people used in studying globular clusters, we find such relaxation
can not be neglect:

$\bullet$ A physical limit exist: one should use at least
about $10^3$ particles to model the smallest halo. If not, the
relaxation process can bring an excessive core collapse within one
Hubble time, especially for the small halos at high redshift.

$\bullet$ Such unwanted core collapse process can leave fossils in the final halos.

$\bullet$ Unfortunately, people used to set $k_{max}=k_{ny}$ in
simulations. It means the smallest halo include only about 8
particles and the physical limit are always broken.

$\bullet$ We give a simple suggestion to avoid such effect. By
setting $k_{max}\leq{k}_*$ but not $k_{ny}$, we can get more
reliable result.

The dynamical properties of CDM particle systems people collect from
N-body simulations have already been basic of many popular
astrophysical topics. Hence we suggest people attach importance to
such a dynamical effect, abating the unwanted fossils and get more
reliable results.


\acknowledgements
We thank Bin Yue, Li Xue, Chang Peng for helpful comments and
discussions on earlier manuscripts.

\end{document}